\documentclass{revtex4-1}
\usepackage{graphics}
\usepackage{hyperref}
\usepackage{amsmath}
\usepackage{graphicx}
\usepackage{amsfonts}
\usepackage{amssymb}
\begin{document}
\title{On the eikonal approach to nuclear diffraction dissociation}
\author{Angela Bonaccorso$^1$ and David M. Brink$^2$ }% etc
%}                     % Do not remove
\affiliation{ $^1$Istituto Nazionale di Fisica Nucleare, Sezione di Pisa, Largo Bruno Pontecorvo 3, 56124 Pisa, Italy.  {\thanks{\emph{Email:} bonac@df.unipi.it}}\\ $^2$Rudolf Peierls Centre of Theoretical Physics, University of Oxford, 1 Keble Road, Oxford OX1 3NP, U.K. }
\begin{abstract}The study of nuclear breakup  of halo and weakly bound particles has been one of the key ingredients in the understanding of exotic nuclei during the last thirty years.
One of the most used methods to analyse data, in particular absolute breakup cross sections, has been the eikonal approximation. Here we revise critically the formalisms used for calculating  the diffraction dissociation part of nuclear breakup and show that there is a formula that can be applied to breakup on any target, while a most commonly used formula must be restricted to light targets as it contains also the effect of Coulomb breakup calculated to first order in the sudden approximation which is  well known for not being  accurate.\end{abstract}%
%\PACS{
 %     {PACS-key}{discribing text of that key}   \and
%      {PACS-key}{discribing text of that key}
%     } % end of PACS codes
%} %end of abstract
%
\maketitle
\section{Introduction}

%TCIDATA{OutputFilter=latex2.dll}
%TCIDATA{CSTFile=article.cst}
%TCIDATA{LastRevised=Thu Apr 05 09:41:39 2001}
%TCIDATA{<META NAME="GraphicsSave" CONTENT="32">}
%TCIDATA{Language=American English}

%%%%%%%%%%%%%%%%%%%%%%%%%%%%%%%%%%%%%%%%%%%%%%%%%%%%%%%%%%%%%%%%%%%%

%%%%%%%%%%%%%%%%%%%%%%%%%%%%%%%%%%%%%%%%%%%%%%%

%%%%%%%%%%%%%%%%%%%%%%%%%%%%%%%%%%%

\vspace{1em}

\begin{figure}


%\resizebox{0.5\textwidth}{!}{%
  \includegraphics [scale=.8,width=8cm,angle=0]{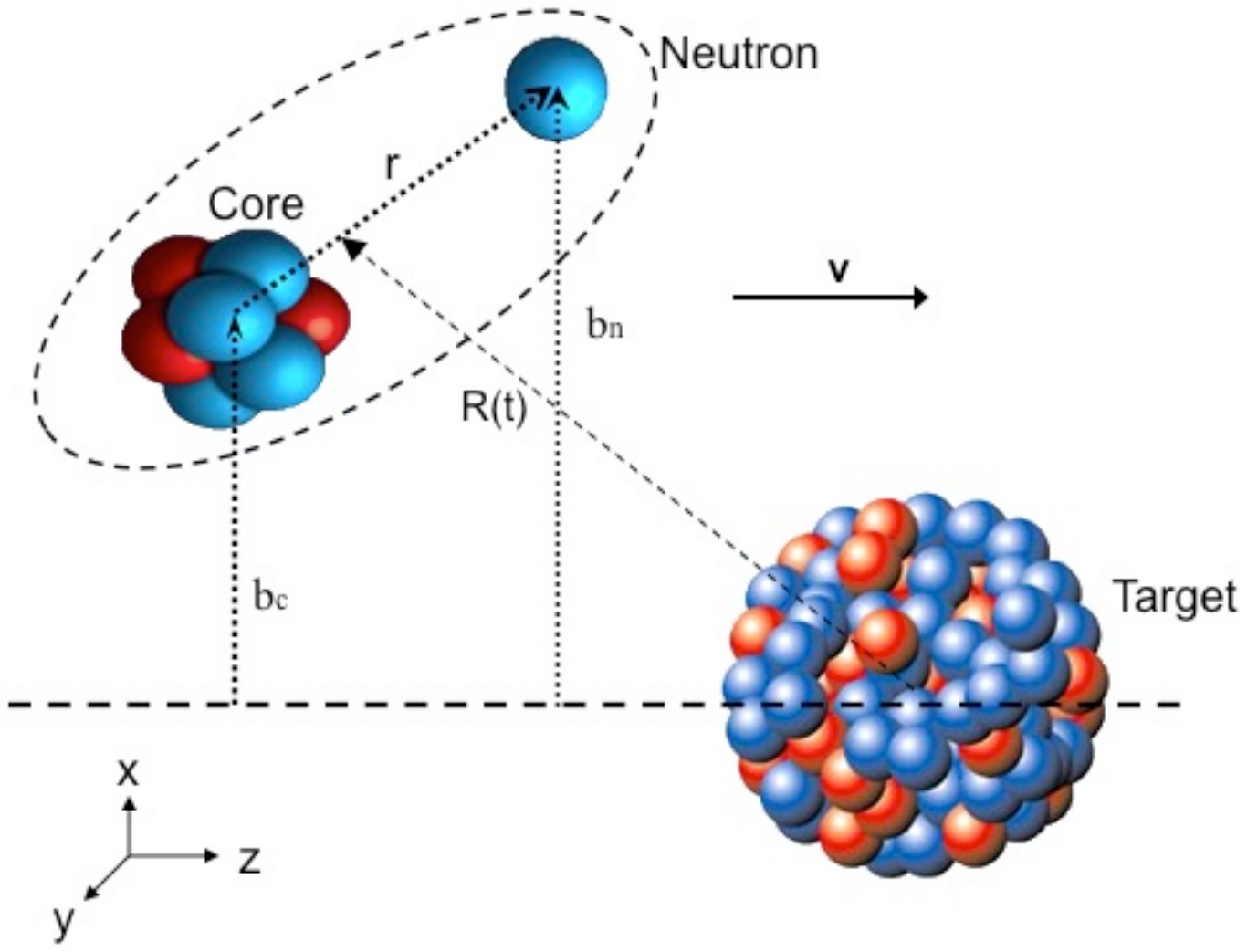}
%}
\caption{Coordinate system.}
%\vspace{5cm}       % Give the correct figure height in cm

\label{fig:1}       % Give a unique label
\end{figure}

The study of nuclear breakup  of halo and weakly bound particles has been one of the key ingredients in the understanding of exotic nuclei during the last thirty years \cite{AB}. Here we discuss and compare formalisms used to calculate the nuclear elastic breakup.
Following Refs. \cite{yaba,esb,BG} we consider a single-particle model for a 
halo nucleus and introduce the eikonal approximation to study its scattering on another target nucleus. The ground state is described by a wave function $\phi
_{0}(\mathbf{r})$ which depends on the relative coordinate \textbf{r} between
the nucleon and the core. After interacting with the target the eikonal
wave-function of the halo nucleus in its rest frame has the form
\begin{equation}
\Psi\left(  \mathbf{r,R}\right)  =S_{n}\left(  \mathbf{b}_{n}\right)  S\left(
\mathbf{b}_{c}\right)  \phi_{0}\left(  \mathbf{r}\right)  \label{e1}%
\end{equation}
where \textbf{R}  and \textbf{r} are the coordinates of the center-of-mass of the projectile
consisting of the core plus one neutron, and of the neutron with respect to the core respectively, see Fig\ref{fig:1}. The vectors
\begin{equation}
\mathbf{b}_{n}=\mathbf{R}_{\perp}+\beta_{2}\mathbf{r}_{\perp}\qquad
\mathrm{and}\qquad\mathbf{b}_{c}=\mathbf{R}_{\perp}-\beta_{1}\mathbf{r}%
_{\perp} \label{e2}%
\end{equation}
are the impact parameters of the neutron and the core with respect to the
target nucleus. Thus $\beta_{1}=m_{n}/m_{p}$, $\beta_{2}=m_{c}/m_{p}%
=1-\beta_{1}$, where $m_{n}$ is the neutron mass, $m_{c}$ is the mass of the
projectile core and m$_{p}=m_{n}+m_{c}$ is the projectile mass. The two
profile functions $S_{n}$ and $S_{c}$ are defined in terms of the
corresponding potentials by
\begin{equation}
S\left(  \mathbf{b}\right)  =\exp\left(  -\frac{i}{\hbar v}\int dzV\left(
\mathbf{b,}z\right)  \right)  \label{e3}%
\end{equation}
where $v$ is the beam velocity. The breakup amplitude generated from the
eikonal wave function (\ref{e1}) has a direct contribution from the
neutron-target optical potential $V_{nT}$ and represented by neutron-target
profile function $S_{n}$ and a core recoil contribution from the core-target
interaction $V_{cT}$ represented by the profile function $S_{c}$ \cite{mbb,mbb2}. The recoil
contribution depends on the ratio $\beta_{1}$ of the neutron mass to the
projectile mass and goes to zero in the limit $\beta_{1}\rightarrow0$. The
potential $V_{cT}$ includes the core-target Coulomb potential and the real and
imaginary parts of the nuclear potential. The Coulomb part of $V_{cT}$ is
responsible for Coulomb breakup. Using the approximate form of  
 the wave function (\ref{e1}) with (\ref{e3}) implies the ''frozen halo'' approximation; the neutron
velocity relative to the core in the projectile and in the final state is slow
compared with the incident velocity $v$.

\bigskip

The eikonal breakup amplitude is defined by \cite{esb}
\begin{widetext}\begin{equation}
A\left(  \mathbf{K,k}\right)  =\int d^{2}\mathbf{R}_{\perp}~e^{-i\mathbf{K_{\perp}\cdot R}%
_{\perp}}\int d^{3}\mathbf{r}~\phi_{\mathbf{k}}^{\ast}\left(  \mathbf{r}%
\right)  \left(  S_{c}\left(  \mathbf{b}_{c}\right)  S_{n}(\mathbf{b}%
_{n})-1\right)  \phi_{0}\left(\mathbf{r}\right).  \label{e4}%
\end{equation}\end{widetext}
The impact parameters $\mathbf{b}_{n}$ and $\mathbf{b}_{c}$ are defined in Eq.(\ref{e2}). The quantities $\left(
\mathbf{K,k}\right)  $ are the momenta conjugate to the coordinates $\left(
\mathbf{R,r}\right).$ They are related to the final momenta of the core,
neutron and target by
\begin{equation}
\mathbf{k}_{c}=-\mathbf{k}+\beta_{2}\mathbf{K},\qquad\mathbf{k}_{n}%
=\mathbf{k}+\beta_{1}\mathbf{K,}\qquad\mathbf{k}_{T}=-\mathbf{K.} \label{e5}%
\end{equation}
The wave function $\phi_{\mathbf{k}}\left(  \mathbf{r}\right) $ is the final
continuum wave function of the neutron relative to the core. The complete
differential cross-section is 
\begin{equation}
\frac{d\sigma}{d^{2}\mathbf{K}d^{3}\mathbf{k}}=\frac{1}{\left(  2\pi\right)
^{5}}\left|  A\left(  \mathbf{K,k}\right)  \right| ^{2} \label{e6}%
\end{equation}
Eq.(\ref{e4}) can also be written as%
\begin{widetext}
\begin{equation}
A\left(  \mathbf{K,k}\right)  =\int d^{2}\mathbf{R}_{\perp}~e^{-i\mathbf{K_{\perp} \cdot R}%
_{\perp}}\int d^{3}\mathbf{r}~\phi_{\mathbf{k}}^{\ast}\left(  \mathbf{r}%
\right)  S_{c}\left(  \mathbf{b}_{c}\right)  S_{n}\left(  \mathbf{b}_{n}\right)\phi
_{0}\left(  \mathbf{r}\right)  \label{e7}%
\end{equation}\end{widetext}
because of the orthogonality of $\phi_{\mathbf{k}}\left(  \mathbf{r}\right) $
and $\phi_{0}\left(  \mathbf{r}\right)  $ (cf. Eq.(8) of Ref.\cite{esb}). This form is
convenient for the developments made in the next paragraph. Equations
(\ref{e7}) is a general eikonal expressions which has been used in
\cite{esb} and by many other authors.

Now we change the integration variable $\mathbf{R}_{\perp}$ in Eq.(\ref{e7})  to
$\mathbf{b}_{c}$ using Eq.(\ref{e2}) and then the amplitude (\ref{e7}) can also be
written as

\begin{widetext}
\begin{equation}
A\left(  \mathbf{K,k}\right)  =\int d^{2}\mathbf{b}_{c}~e^{-i\mathbf{K_{\perp} \cdot b}_{c}%
}S_{c}\left(  \mathbf{b}_{c}\right) \int d^{3}\mathbf{r}~\phi_{\mathbf{k}%
}^{\ast}\left(  \mathbf{r}\right)  e^{\left(  -i\beta_{1}\mathbf{K}_{\perp
}\cdot \mathbf{r}_{\perp}\right)  }S_{n}\left(  \mathbf{b}_{n}\right)  \phi
_{0}\left(  \mathbf{r}\right)  \label{e8}%
\end{equation}\end{widetext}
where $\mathbf{b}_{n}=\mathbf{b}_{c}+\mathbf{r}_{\perp}$. The scattering
amplitude (\ref{e8}) is a full 3-body eikonal amplitude. It is exactly
equivalent to (\ref{e4}) and (\ref{e7}). \ Both $\mathbf{K}$ and $\mathbf{k}$
are observables which can, in principle, be measured. The next step is to
write Eq.(\ref{e8}) as%
\begin{equation}
A\left(  \mathbf{K,k}\right)  =\int d^{2}\mathbf{b}_{c}~e^{-i\mathbf{K_{\perp} \cdot b}_{c}%
}S_{c}\left(  \mathbf{b}_{c}\right) \int d^{3}\mathbf{r}~\phi_{\mathbf{k}%
}^{\ast}\left(  \mathbf{r}\right)  (e^{\left(  -i\beta_{1}\mathbf{K}_{\perp
}\cdot \mathbf{r}_{\perp}\right)  }S_{n}\left(  \mathbf{b}_{n}\right)  -1)\phi
_{0}\left(\mathbf{r}\right), \nonumber \\
\label{e9}%
\end{equation}
where we have again used the orthogonality of $\phi_{\mathbf{k}}\left(
\mathbf{r}\right)  $ and $\phi_{0}\left(  \mathbf{r}\right)  $. Provided that there are no
core neutron resonances in the final state we can make the  approximation to neglect the final state interaction of the neutron with the
projectile core and replace
the final continuum state $\phi_{\mathbf{k}}\left(
\mathbf{r}\right)  $ in Eq.(\ref{e9}) by a plane wave $e^{
i\mathbf{k\cdot r}}  $. \ Then we obtain $\bf {k}$ from the definition of  ${\bf {k}}_n$ in Eq. (\ref{e5})
and the transverse
component of the final neutron momentum is $\mathbf{k}_{n\perp}=\mathbf{k}%
_{\perp}+\beta_{1}\mathbf{K}_{\perp}.$\textbf{\ }

 Thus the amplitude (\ref{e9}) becomes
 
 \begin{equation}
A\left(  \mathbf{K,k}\right)  =\int d^{2}\mathbf{b}_{c}~e^{-i\mathbf{K_{\perp} \cdot b}_{c}%
}S_{c}\left(  \mathbf{b}_{c}\right)  \int  d^{3}\mathbf{r}~e^{-i(\mathbf{k}_{n}-\beta_{1}\mathbf{K})%
\cdot \mathbf{r}} (e^{\left(  -i\beta_{1}\mathbf{K}_{\perp
}\cdot \mathbf{r}_{\perp}\right)  }S_{n}\left(  \mathbf{b}_{n}\right)  -1 )   \phi
_{0}\left(  \mathbf{r}\right). \label{ea9}%
\end{equation}
which  can be further  written as
\begin{equation}
A\left(  \mathbf{K,k}\right)  =-\int d^{2}\mathbf{b}_{c}~e^{-i\mathbf{K_{\perp} \cdot b}_{c}%
}S_{c}\left(  \mathbf{b}_{c}\right)  g(\mathbf{k}_{n}\mathbf{b}_{c}),
\label{eb1}%
\end{equation}
where%

\begin{equation}
g(\mathbf{k}_{n}\mathbf{,b}_{c})\approx\int d^{3}\mathbf{r}~e^{-i\mathbf{k}_{n}\cdot \mathbf{r}}%
\left(  e^{\left(  i\beta_{1}\mathbf{K}_{\perp}\cdot \mathbf{r}_{\perp
}\right)  }-S_{n}\left(  \mathbf{b}_{n}\right)  \right)  \phi_{0}\left(
\mathbf{r}\right).  \label{eb2}%
\end{equation}
Then we write the total breakup amplitude as a sum
\begin{equation}
g(\mathbf{k}_{n}\mathbf{,b}_{c})=g_{n}(\mathbf{k}_{n}\mathbf{,b}_{c}%
)+g_{c}(\mathbf{k}_{n}\mathbf{,b}_{c})\label{eb3}%
\end{equation}
where
\begin{widetext}
\begin{align}
g_{n}(\mathbf{k}_{n}\mathbf{,b}_{c}) &  =\int d^{2}\mathbf{r}_{\perp
}~e^{-i\mathbf{k}_{n}\cdot\mathbf{r}_{\perp}}\left(  1-S_{n}\left(  \mathbf{b}%
_{n}\right)  \right) {\tilde \phi}_{0}\left(  \mathbf{r}_{\perp},k_{z}\right)
\label{eb4}\\
 g_{c}(\mathbf{k}_{n},\mathbf{K}_{\perp}\mathbf{,b}_{c}) &  =\int
d^{2}\mathbf{r}_{\perp}~e^{-i\mathbf{k}_{n}\cdot \mathbf{r}_{\perp}}\left(
e^{\left(  i\beta_{1}\mathbf{K}_{\perp} \cdot \mathbf{r}_{\perp}\right)  }-1\right)
{\tilde \phi}_{0}\left(  \mathbf{r}_{\perp},k_{z}\right).  \label{eb5}%
\end{align}
\end{widetext}
Here $k_{z}$ is the z-component of the final neutron momentum and $\tilde \phi
_{0}\left(  \mathbf{r}_{\perp},k_{z}\right)  $ is the one-dimensional Fourier
transform of the initial wave function with respect to the z-coordinate. The
amplitude $g_{n}$ is just the eikonal amplitude used in Ref. \cite{mbb}. It depends on
the target neutron interaction through the profile function $S_{n}\left(
\mathbf{b}_{n}\right)  $. The second integral $g_{c}$ is the recoil breakup
amplitude. It depends on the recoil momentum $\mathbf{K}_{\perp}$. %(Just for
%us: $g_{c}$ is a kind of form-factor depending on the momentum transfer, just
%like in Ron Johnson's talk). 

In the following paragraph we simplify Eq.(\ref{eb5}) by making a
semi-classical approximation. If the core-target profile function
$S_{c}\left(  \mathbf{b}_{c}\right)  $ is smooth and $K_{\perp}$ is large
enough $\left(  K_{\perp}b_{c}>>1\right)  $ then the integral over
$\mathbf{b}_{c}$ in (\ref{eb1}) can be estimated by the method of stationary
phase. The dominant contribution comes from $\mathbf{K}_{\perp}$ parallel to
$\mathbf{b}_{c}$ and at the stationary point it will be approximated by the
classical momentum transfer
\begin{equation}
\mathbf{K}_{\perp}\approx\mathcal{K}_{\perp}\left(  \mathbf{b}_{c}\right)
=\frac{1}{\hbar}\int\mathbf{F}_{cT}\left(  \mathbf{b}_{c},vt\right)
dt\label{eb6}%
\end{equation}
where $\mathbf{F}_{cT}=-\nabla V_{cT}$ is the classical force on the
projectile core due to the core-target interaction and the integral is
calculated along the path with impact parameter $\mathbf{b}_{c}$. For a full
semi-classical evaluation of the $\mathbf{b}_{c}$ integral in Eq.(\ref{eb1})
we have to assume that for each value of $\mathbf{K}_{\perp}$ there is a
unique \ core-target impact parameter which satisfies Eq.(\ref{eb6}). At this
stage we do not do this but instead approximate $\mathbf{K}_{\perp}$ in the
integral in Eq.(\ref{eb5}) by its semi-classsical value. For each value of
$\mathbf{b}_{c}$ there is a unique $\mathcal{K}_{\perp}\left(  \mathbf{b}%
_{c}\right)  $ given by Eq.(\ref{eb6}). This approximation results in a
decoupling of the two integrals in (\ref{eb1}) where now the recoil amplitude
\begin{equation}
g_{c}(\mathbf{k}_{n}\mathbf{,b}_{c})=\int d^{2}\mathbf{r}_{\perp
}~e^{-i\mathbf{k}_{n}\cdot \mathbf{r}_{\perp}}\left(  e^{\left(  i\beta
_{1}\mathcal{K}_{\perp}\left(  \mathbf{b}_{c}\right)  \cdot \mathbf{r}_{\perp
}\right)  }-1\right)  \phi_{0}\left(  \mathbf{r}_{\perp},k_{z}\right)
\label{eb8}%
\end{equation}
is a function of $\mathbf{k}_{n}$ and $\mathbf{b}_{c}.$

With this approximation the breakup cross-section as a function of the neutron
momentum $\mathbf{k}_{n}$ when $\mathbf{K}_{\perp}$ is not observed is
\begin{equation}
\frac{d\sigma}{d^{3}\mathbf{k}_{n}}=\int\left|  A\left(  \mathbf{K,k}\right)
\right|  ^{2}d^{2}\mathbf{K}_{\perp}=\int d^{2}\mathbf{b}_{c}~P_{el}\left(
b_{c}\right)  \left|  g(\mathbf{k}_{n}\mathbf{,b}_{c})\right|  ^{2}.
\label{ek3}%
\end{equation}
Here $P_{el}\left(  b_{c}\right)  =\left|  S_{c}\left(  \mathbf{b}_{c}\right)
\right|  ^{2}$ is the probability that the core remains in its ground state
during the collision. 
Different ways of calculating it and their respective accuracies have been recently revised in Ref.\cite{bcc}.

%(If we make a full stationary phase approximation to the integral over
%$\mathbf{b}_{c}$ in (\ref{e8}) the the cross-section factorizes
%\begin{equation}
%\frac{d\sigma}{d^{2}\mathbf{K}d^{3}\mathbf{k}}=\left(  \frac{d\sigma}%
%{d^{2}\mathbf{K}}\right)  _{class}\left|  g(\mathbf{k}_{n}\mathbf{,b}%
%_{c})\right|  ^{2} \label{ek5}%
%\end{equation}
%Here the term $\left(  d\sigma/d^{2}\mathbf{K}\right)  _{class}$ is the
%classical core-target elastic scattering cross-section.)

 %The form (\ref{ek5})
%should be a good approximation for values of $\mathbf{K}$ at which the
%core-target Coulomb interaction is dominant. In this limit the reaction plane
%defined by $\mathbf{K}$ (or $\mathbf{k}_{n}+\mathbf{k}_{c}$) and the z-axis is
%parallel to the plane defined by the z-axis and $\mathbf{b}_{c}$.

The Coulomb breakup of an odd-neutron nucleus like $^{11}$Be is due to the
core-target interaction. Its contribution is included in the recoil amplitude
(\ref{eb5}) or (\ref{eb8}).When the recoil effect is small enough the
exponential factor in Eq.(\ref{eb8}) can be expanded to first order in
$\beta_{1}$ and the recoil amplitude reduces to the standard dipole form in
the eikonal limit%
\begin{equation}
g_{c}(\mathbf{k}_{n}\mathbf{,b}_{c})=i\beta_{1}\int d^{2}\mathbf{r}_{\perp
}~e^{-i\mathbf{k}_{n}\cdot \mathbf{r}_{\perp}}~\mathcal{K}_{\perp}\left(
\mathbf{b}_{c}\right)  \cdot \mathbf{r}_{\perp}~{\tilde \phi}_{0}\left(  \mathbf{r}_{\perp
},k_{z}\right).
\label{20}\end{equation}
The explicit expression for the momentum transfer (\ref{eb6}) is
\begin{equation}
\mathcal{K}_{\perp}\left(  \mathbf{b}_{c}\right)  =\frac{2Z_{P}Z_{T}e^{2}%
}{\hbar vb_{c}^{2}}\mathbf{b}_{c}.%
\end{equation}
Choosing the x-axis in the direction of $\mathbf{b}_{c}$, Eq.(\ref{20}) reduces to%
\begin{equation}
g_{c}(\mathbf{k}_{n}\mathbf{,b}_{c})=\beta_{1}\frac{2Z_{P}Z_{T}e^{2}}{\hbar
vb_{c}}\frac{\partial}{\partial k_{x}}\tilde{\phi}_{0}\left(
\mathbf{k}\right).\label{22}
\end{equation}
The Coulomb amplitude calculated in the standard dipole  approximation by time dependent perturbation theory \cite{mbb} reads%
\begin{widetext}
\begin{equation}
g_{c}(\mathbf{k,b}_{c}) =  \beta_{1}\frac{2Z_{P}Z_{T}e^{2}}{\hbar vb_{c}%Á
}\left(  \bar{\omega}K_{1}\left(  \bar{\omega}\right)  \frac{\partial
}{\partial k_{x}}+i\bar{\omega}K_{0}\left(  \bar{\omega}\right)
\frac{\partial}{\partial k_{z}}\right)  \tilde{\phi}_0\left(\mathbf{k}\right).%
\end{equation}\end{widetext}

Thus Eq.(\ref{22}) is just the sudden limit of the usual dipole Born approximation.
Infact when the adiabaticity parameter%
\begin{equation}
\bar{\omega}=\frac{\varepsilon_{k}-\varepsilon_{0}}{\hbar v}b_{c}%
\end{equation}
is small  the sudden limit  $\bar \omega\to 0$ applies and $ \bar{\omega}K_{1}\left(  \bar{\omega}\right) \to 1$ and  $\bar{\omega}K_{0}\left(  \bar{\omega}\right) \to 0$.

%\bigskip\lbrack Some numbers for $^{11}$Be at $E_{lab}=70$ MeV/A and for a
%$^{208}$Pb target: $\varepsilon_{k}-\varepsilon_{0}\approx4$ MeV, $v/c=0.39$,
%$\hbar v=76.8$ fm. $b_{c}\approx R_{s}\approx11.2$ fm. Hence $\bar{\omega}%
%\sim0.57$. Put $Z_{P}Z_{T}e^{2}/\hbar v=C=6.1$, $\beta_{1}\mathcal{K}_{\perp
%\left(  \mathbf{b}_{c}\right)  .\mathbf{r}_{\perp}\sim2\beta_{1}C\left(
%R_{0}/R_{s}\right)  \sim0.4$.\rbrack
%\vskip 1pt

%Equation \ref{ek3} is what we  have used previously and we have been able to calculate both the nuclear and Coulomb amplitudes consistently \cite{mbb}. 

If we consider only the nuclear part of the amplitude Eq.(\ref{eb4}) and integrate over the neutron momentum $d^3{\bf k}_n$ the total  nuclear diffraction 
cross section becomes \cite{mbb,mbb2}

\begin{equation}
\sigma_{-n}=\int d^{2}\mathbf{b}_{c}~|S_{ct}\left(\mathbf
{b}_{c}\right) |^2\int d^{2}\mathbf{r}_{\perp}~| 1-S_{n}\left(  \mathbf{b}%
_{n} \right|^2 |{\tilde \phi}_{0}\left(  \mathbf{r}_{\perp}\right)|^2.
\label{ek33}%
\end{equation}
The general eikonal expressions Eq.
(\ref{e7}) and (\ref{eb2})  have been used instead in
\cite{yaba,esb,BG} and by many other authors, without the steps discussed above to separate the nuclear and Coulomb parts thus  leading to Eq.(2.19) of \cite{yaba}, Eq.(10) of \cite{esb} and  Eq.(13) of \cite{BG} which are all equivalent and read
\begin{equation}
\sigma_{-n}=\int d^{2}\mathbf{b}_{c}\int d^{3}\mathbf{r}~| S_{ct}\left(
\mathbf{b}_{c}) S_{n}\left(  \mathbf{b}%
_{n}\right ) \right|^2 |{ \phi}_{0}\left(  \mathbf{r}\right)|^2-  \int  d^{2}\mathbf{b}_{c} ~\left| \int d^{3}\mathbf{r}~S_{ct}(
\mathbf{b}_{c}) S_{n}\left(  \mathbf{b}%
_{n}\right ) \ |{ \phi}_{0}\left(  \mathbf{r}\right)|^2\right |^2.
\label{ek44}%
\end{equation}

 On the basis of the previous equations  if one goes from Eq.(\ref{e7})  to Eq.(\ref{ek3}) and then  to Eq. (\ref{ek44}) the effect of recoil is automatically included, which corresponds to what is also usually called Coulomb breakup calculated in the sudden limit. It is well known that the sudden approximation to the Coulomb breakup gives too large cross sections \cite{mbb,mbb2,BT}. Therefore when applied to breakup on a heavy target results  from Eq.(\ref{ek44})  will be larger than what one would get from Eq.(\ref{ek33}) where only the nuclear part of the amplitude has been used. % In fact looking at Fig. (2a) of Ref.\cite{esb} for a heavy target (A=197) one can see that the calculated nuclear breakup cross section is larger than the Coulomb cross section which is not easy to understand. 

We conclude therefore that Eq.(\ref{ek44}) cannot be defined as the equation representing  only the nuclear elastic breakup part. From the point of view of the formalism it contains also the Coulomb breakup calculated in the sudden approximation. On light targets it  gives  close results to Eq.(\ref{ek33}), within other numerical incertitudes. On the other hand on heavy targets it definitely gives too large cross sections, also because it contains an interference term and it would be particularly unreliable to make predictions on the nuclear part of elastic breakup while  Eq.(\ref{ek33}) provides a safer way.  Finally we propose to follow the method introduced in \cite{mbb} and \cite{mbb2} to calculate consistently nuclear and Coulomb breakup for a neutron, while for proton breakup we propose to follow \cite{alv2,rav}. In these references Coulomb breakup is treated to all orders and all multipolarities.

Comparisons between numerical results of Eqs.(\ref{ek33},\ref{ek44}) will be presented elsewhere.



%

% The section below may be edited at your convenience to acknowledge 
% each author's contribution to the manuscript.
% You may remove it if you are a single author.
%
\bigskip
We thank Ravinder Kumar for drawing the figure.

%
% BibTeX users please use
% \bibliographystyle{}
% \bibliography{}

\end{document}